\PassOptionsToPackage{usenames,dvipsnames}{xcolor}
\documentclass[sigconf, nonacm]{acmart}
\usepackage{balance}  
\usepackage{mathtools}
\usepackage[shortcuts]{extdash}
\usepackage[utf8]{inputenc}
\usepackage{csquotes}
\usepackage{IEEEtrantools}
\usepackage{textcomp}
\usepackage{lstautogobble}
\usepackage{needspace}
\usepackage{suffix}
\usepackage{hyperref}
\usepackage{zi4}
\usepackage[font=normal,skip=4pt]{caption}
\usepackage{xspace}
\usepackage{graphicx} 
\usepackage{xcolor}
\usepackage{enumitem}
\graphicspath{{images/}} 
\AtBeginDocument{%
  \providecommand\BibTeX{{%
    \normalfont B\kern-0.5em{\scshape i\kern-0.25em b}\kern-0.8em\TeX}}}

\newcommand\vldbdoi{10.14778/3554821.3554871}
\newcommand\vldbpages{3670-3673}
\newcommand\vldbvolume{15}
\newcommand\vldbissue{12}
\newcommand\vldbyear{2022}
\newcommand\vldbauthors{\authors}
\newcommand\vldbtitle{\shorttitle}
\newcommand\vldbavailabilityurl{} 
\newcommand\vldbpagestyle{empty}

\newcommand{\screenshot}[1]{\includegraphics[width=0.97\linewidth]{#1}}
\WithSuffix\newcommand\screenshot*[1]{\includegraphics[width=1.0\textwidth]{#1}}

\newcommand{\sql}[1]{\texttt{#1}}

\newcommand{\sling}[1]{\texttt{#1}}

\WithSuffix\newcommand\subparagraph*[1]{\par \noindent \sloppy \textbf{#1.}}

\newcommand{\ignore}[1]{}
\newcommand{\subhead}[1]{\vspace{0.3\baselineskip}\noindent\textbf{#1}}

\newcommand{\code}[1]{\texttt{#1}}
\definecolor{tomato}{rgb}{1,0.2,0}
\definecolor{turqoise}{rgb}{0.03, 0.91, 0.87}
\definecolor{grey}{rgb}{0.4,0.4,0.4}
\newif\ifnotes
\notesfalse
\newcommand{\cagatay}[1]{\ifnotes{\small[\textcolor{grey}{\c{C}a\u{g}atay:}\textcolor{tomato}{#1}]}\fi}
\newcommand{\jlg}[1]{\ifnotes{\small[\textcolor{grey}{jlg:}\textcolor{tomato}{#1}]}\fi}

\setcopyright{none}
\makeatletter
\def\@copyrightspace{\relax}
\makeatother

\begin{document}

\title{Sigma Workbook: A Spreadsheet for Cloud Data Warehouses }

\author{James Gale}
\affiliation{%
  \institution{Sigma Computing}
}
\email{jlg@sigmacomputing.com}

\author{Max Seiden}
\affiliation{%
  \institution{Sigma Computing}
}
\email{max@sigmacomputing.com}

\author{Deepanshu Utkarsh}
\affiliation{%
  \institution{Sigma Computing}
}
\email{deepanshu@sigmacomputing.com}

\author{Jason Frantz}
\affiliation{%
    \institution{Sigma Computing}
}
\email{jason@sigmacomputing.com}

\author{Rob Woollen}
\affiliation{%
    \institution{Sigma Computing}
}
\email{rwoollen@sigmacomputing.com}

\author{\c{C}a\u{g}atay Demiralp}
\affiliation{%
  \institution{Sigma Computing}
}
\email{cagatay@sigmacomputing.com}

\renewcommand{\shortauthors}{Gale and Seiden, et al.}



\begin{abstract}
Cloud data warehouses (CDWs) bring large-scale data and compute power closer to users in enterprises. However, existing tools for analyzing data in CDWs are either limited in ad-hoc transformations or difficult to use for business users. Here we introduce Sigma Workbook, a new interactive system that enables business users to easily perform visual analysis of data in CDWs at scale. For this, Sigma Workbook provides an accessible spreadsheet-like interface for analysis through direct manipulation. Sigma Workbook dynamically constructs matching SQL queries from user interactions, building on the versatility and expressivity of SQL. Constructed queries are directly executed on CDWs, leveraging the superior characteristics of the new generation CDWs, including scalability. We demonstrate Sigma Workbook through 3 real-life use cases---cohort analysis, sessionization, and data augmentation---and underline Workbook's ease of use, scalability, and expressivity. 
\end{abstract}
\settopmatter{printacmref=false, printfolios=true,printccs=false}
\renewcommand\footnotetextcopyrightpermission[1]{}

\settopmatter{printacmref=false}

\begin{teaserfigure}
  \centering
  \includegraphics[width=0.99\textwidth]{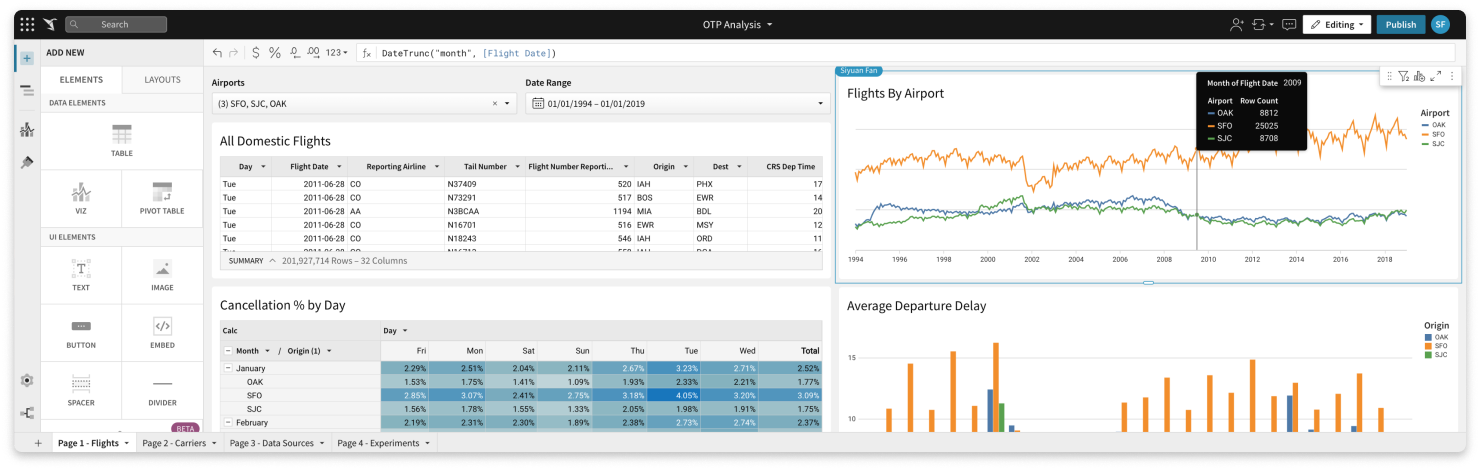}
  \caption{Sigma Workbook is an interactive workspace for analyzing enterprise-scale data in cloud data warehouses (CDWs). Its interface builds on spreadsheets while automatically compiling data operations to SQL queries and executing them on the CDW. Workbook  enables users to benefit from the characteristics of SQL and CDWs using their knowledge of spreadsheets.\label{fig:teaser}}
  \Description{A screenshot of the Sigma Workbook interface.}
\end{teaserfigure}

\maketitle

\pagestyle{\vldbpagestyle}
\begingroup\small\noindent\raggedright\textbf{PVLDB Reference Format:}\\
\vldbauthors. \vldbtitle. PVLDB, \vldbvolume(\vldbissue): \vldbpages, \vldbyear.\\
\href{https://doi.org/\vldbdoi}{doi:\vldbdoi}
\endgroup
\begingroup
\renewcommand\thefootnote{}\footnote{\noindent
This work is licensed under the Creative Commons BY-NC-ND 4.0 International License. Visit \url{https://creativecommons.org/licenses/by-nc-nd/4.0/} to view a copy of this license. For any use beyond those covered by this license, obtain permission by emailing \href{mailto:info@vldb.org}{info@vldb.org}. Copyright is held by the owner/author(s). Publication rights licensed to the VLDB Endowment. \\
\raggedright Proceedings of the VLDB Endowment, Vol. \vldbvolume, No. \vldbissue\ %
ISSN 2150-8097. \\
\href{https://doi.org/\vldbdoi}{doi:\vldbdoi} \\
}\addtocounter{footnote}{-1}\endgroup

\ifdefempty{\vldbavailabilityurl}{}{
\vspace{.3cm}
\begingroup\small\noindent\raggedright\textbf{PVLDB Artifact Availability:}\\
The source code, data, and/or other artifacts have been made available at \url{\vldbavailabilityurl}.
\endgroup
}

\section{Introduction}\label{sec:intro}
Enterprise data is increasingly stored in cloud data warehouses (CDWs) as they enable the storage of large-scale datasets with reliability and compliance guarantees while reducing costs. Business users (non-technical domain experts such as operations associates, marketing managers, and product managers) in enterprises wish to use this data for augmenting their decision-making, ideally without going through analysts. Many of these users are comfortable doing analyses using spreadsheets. However, due to their limited scalability and expressivity~\cite{bendre2018towards,rahman2020benchmarking}, existing spreadsheet applications aren’t adequate for accessing and analyzing data residing within CDWs.

In this paper, we introduce Sigma Workbook\footnote{A demo video of Sigma Workbook is available at:
\url{https://tinyurl.com/sigma-workbook}.} (Workbook for short), a SaaS system (Figure~\ref{fig:teaser}) that enables business users to perform interactive ad-hoc analysis on datasets stored in CDWs. It aims to effectively combine ease of use, expressivity, and scalability in order to support iterative visual data analysis of enterprise data.  To enhance accessibility, Workbook integrates an easy-to-use, intuitive
interface with affordances that have made spreadsheet applications successful~\cite{nardi1990}, including a simple
expression language embedded in a table of values, easy
references, easy refactoring, and isolation of errors. Workbook dynamically compiles data operations interactively specified through this interface into SQL queries, building on the versatility and expressivity of SQL. This amplifies users' ability to generate complex queries that can be otherwise daunting to manually specify. 

Sigma Workbook executes compiled queries on CDWs, directly leveraging their desirable properties such as scalability, security (e.g., compliance with regulations such as HIPPA and GDPR), elasticity, and reliability~\cite{gupta2015amazon, Dageville:2016}. This direct, interactive interface to the CDW differentiates Workbook from the architecture of current BI systems, which can produce beautiful visualizations and dashboards but then be limited when users want to get to the ``row-level'' data behind their dashboards. Unlike spreadsheet applications, Sigma Workbook enables users to explore billions of records or terabytes of data. In this sense, Sigma Workbook is an accumulation of the decades-long ideas proposing to combine the accessibility of spreadsheet-like direct manipulation with the characteristics of database systems that enterprises rely on ~\cite{liu2009spreadsheet,raman1999scalable,witkowski2003spreadsheets,tyszkiewicz2010spreadsheet,bakke2011spreadsheet,bendre2015dataspread}.

We demonstrate Workbook here with 3 use cases:
cohort analysis, sessionization, and augmentation with user-created data. We use a dataset of 200M flight records for all the use cases.

\begin{figure}[tbp]
    \screenshot{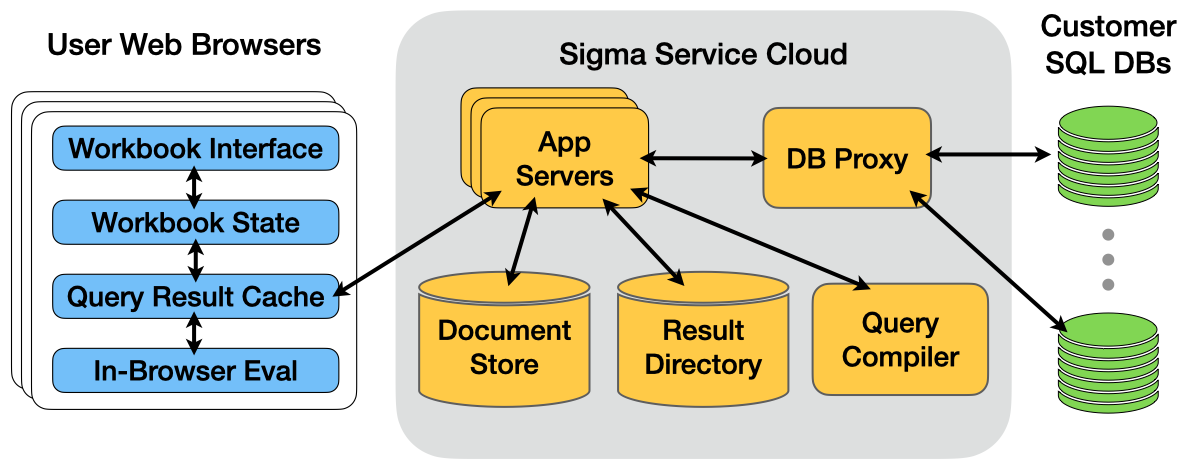}
  \caption{Sigma Workbook architecture. Web browsers running the Sigma
application use Sigma's multi-tenant cloud service to interface
with the user's own CDW. The app server acts as an intermediary for
query requests that are compiled to SQL and a proxy performs workload
scheduling and interfacing with the database. Query results
are returned to the result cache in the browser app and presented to the user.
 \label{fig:architecture}}
\end{figure}

\begin{figure*}[tbh]
  \screenshot*{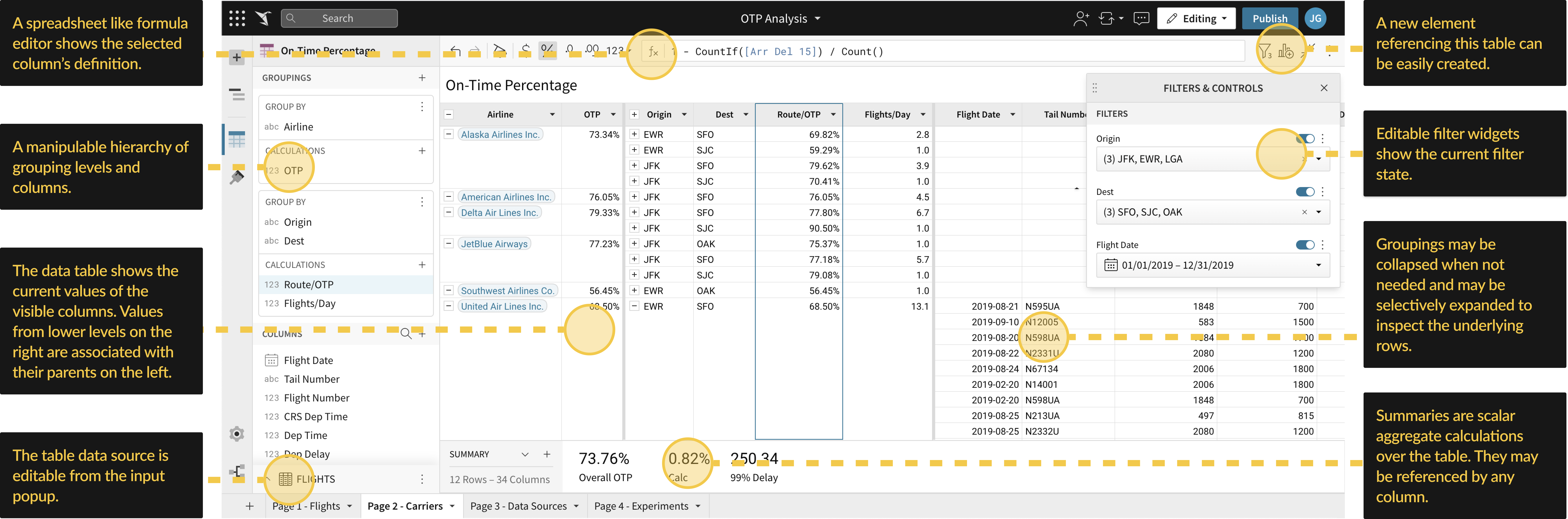}
  \caption{The Workbook table element (annotated) is a query defined by grouping levels, columns, and filters.\label{ss:table}}
  \ignore{\caption{The central feature
    is the data table (E), which displays the results of the query
    described by the worksheet. \cagatay{Maybe, rephrase the following for clarity?} Here a portion of the base level is
    collapsed except a single grouping (F), expanded for viewing. The grouping inspector (A) summarizes the
    multi-dimensional hierarchy of aggregation in the table and shows the schema (D) of the table. The
    formula bar (B) enables the inspection of calculated value
    definitions. The control panel (C) displays the filter and
    state and the table summary (H) gives the
    scale of the data table and other scalar calculations. The table source
    is editable from the source pop-up (G).
    Each of these elements supports editing
    through direct manipulation.\label{ss:table}}}
\end{figure*}

\section{System Overview}\label{sec:overview}
  Workbook is the unified interface for exploring and presenting
data in Sigma, a multi-tenant Software-as-a-Service (SaaS) web application
for BI. The elements of the Workbook implementation
(Figure~\ref{fig:architecture}) support the interactive construction,
composition and visualization of SQL queries through direct
manipulation.

    Sigma customers configure the service
    with access to a CDW they control. No pre-processing or
    ingestion is required before users can begin OLAP through Sigma. Sigma allows multiple warehouse configurations per customer, currently
    supporting Data\-bricks~\cite{photon}, Big\-Query~\cite{bigquery,melnik2010dremel},
    PostgreSQL~\cite{postgresql}, Redshift~\cite{redshift} and
    Snowflake~\cite{snowflake,Dageville:2016}. The elastic and scalable
    nature of CDWs enables them to support diverse, interactive OLAP
    workloads from large numbers of users. Sigma benefits from these
    desirable characteristics by pushing computation to CDWs.

    The Workbook interface enables users to find and reference the tables within their database to construct new analyses (Section \ref{sec:interface}).
    Workbook state can be saved and restored as a document. These
    documents can be named and organized in a file system within Sigma and
    may be shared or copied. Unnamed Workbook documents are stored as
    persistent, anonymous ``explorations'' which can be easily discarded.
    
    Access to the customer's data warehouse by the Sigma web application is always mediated by the Sigma service. Interactive data operations
    expressed by a user are sent to the Sigma service as a JSON-encoding
    of the Workbook state. The Sigma service performs
    authentication, access control checks, query input graph resolution, and
    materialized view substitution. The validated, fully resolved query graph is
    compiled into a corresponding SQL query. The
    SQL query is then placed into a workload management queue and
    subsequently executed in the customer's database. The query results are fetched from the database and forwarded directly back to the web application for presentation.

\section{Workbook Interface}\label{sec:interface}
The Workbook interface is designed with accessibility in mind, 
incorporating successful features of spreadsheets to enable
business users to easily explore and analyze tables in 
relational databases. It provides a canvas where a user can add and
arrange elements of different types. There are 3 categories of Workbook elements: (1)~data elements, including tables, visualizations, pivot tables, and user-created tables; (2)~user interface elements, including text, images and spacers; and (3)~interactive control elements, including sliders, lists, text inputs, date pickers, and drill-downs. Users can partition the canvas into 
pages to organize their analysis.

\subsection{Tables}
The Workbook table element~(Figure \ref{ss:table}) is an evolution of our
earlier work~\cite{gale2021sigma} enabling users to interactively construct 
and manipulate database queries. Workbook tables are defined by 3 constructs: 
(1)~the grouping levels of the table; (2)~the formula definitions 
for columns in the table; and (3)~zero or more filters applied 
to the table. The specification also describes the input data sources 
and various data formatting options.

\subhead{Grouping Levels}
To perform aggregation and window calculations, users need a
mechanism to specify grouping and sorting in the Workbook. 
In a Workbook table, users define a
list of grouping levels (level for short) that visually arrange
records in a nested fashion. A level specifies a grouping key,
map of columns, and an ordering annotation---this enables
expressions such as \sling{CountDistinct} or \sling{MovingAverage} to
derive grouping and ordering properties.


Levels in the Workbook table are organized in a hierarchy. The table
always has at least two levels. The lowest level is known as the
\textit{base}, and initially contains only columns that reference the
input data source. It is the only level that does not have keys 
and is not aggregated. The highest level, known as the
\textit{summary}, has an empty key set and is used to calculate scalar
aggregates. There can be zero or more levels between these two. The
only restriction is that level keys must reference
columns from a lower level.


\subhead{Columns}
Each table column is defined by an expression, its
visibility, and its ``resident level.'' Column expressions, known as
formulas, are written in an expression language familiar to users of
spreadsheet and BI tools. Like SQL, supported functions
fall into one of three categories: single row, aggregate, and
window. However, much like a spreadsheet, there are no restrictions 
on how these functions are composed.


\subhead{Filters}
The Workbook table provides specialized filter widgets that apply a
predicate to a column's values to select records from the
data table. As the table does not allow users to specify an explicit order
of operations, Workbook implements a behavior that is explainable and
predictable: filters are applied in a greedy manner, as soon as their dependencies are met.  

\subhead{Data Sources}
Every Workbook table has a data source, being either a
database table, a SQL query, an uploaded CSV file, or another Workbook
data element. Additional inputs can be included from the same types of
sources via joins or unions.

\vspace{-5pt}
\subsection{Ad-hoc Joins with Lookup and Rollup}
Column formulas can use a special function, \sling{Lookup}, to pull-in
values from other elements in a manner similar to the common
spreadsheet function, VLOOKUP. \sling{Lookup} expressions behave
like a foreign-key left-join in that they never affect the cardinality
of the query. 

\sling{Rollup} is similar to \sling{Lookup}, and takes an
aggregate expression to be evaluated against the join target. In fact,
\sling{Lookup} is a special case of \sling{Rollup} with the virtual aggregate
ATTR wrapping the scalar expression. Both \sling{Lookup} and
\sling{Rollup} support self-joins.

These formulas can be input directly by the user, but Workbook also
includes guided interfaces to help users construct and re-use these
relationships between elements.

\subsection{Visualizations and Pivot Tables}

Workbook visualization elements use Vega~\cite{satyanarayan2016vega} and support common visualization types. Pivot tables can also be defined. Like tables, visualization and pivot table elements include columns and filters.
Similarly, both elements have a data source and may be a source for other elements.
\subsection{Ad-hoc Data}
Workbook supports directly adding and updating data, enabling users to augment their analysis with provisional data and run what-if scenarios.
Workbook users can create free-form, editable tables, which are projected into the warehouse. When the values in these tables are changed (e.g., by editing in values or copy-and-pasting from a spreadsheet), the edits are propagated to the warehouse. 
Users can also add their own CSV data as sources to any workbook element. The parsed file is transparently marshaled into the user's warehouse as a database table. 

\subsection{Presentation and Collaboration}
Workbook supports collaborative analysis. Editing of a Workbook document is ``multi-player,'' allowing editors to observe each other's changes in real-time.
User can create comments on elements in a document and view the history of edits. 
Workbook supports a viewer mode tailored for presentations while allowing some limited, transient exploration. Workbook also supports live document embedding into other websites.

\subhead{Layout} Workbooks are edited in a ``desktop'' web browser but
may be viewed in many form factors, including mobile devices. Workbook
elements are laid out as a sequence of sections, each divided into a
number of columns, similar to interactive website builders. The
layout is responsive to varying screen sizes so
that the elements are still legible and interactable. 
Workbook offers the inclusion of presentation elements, such as images, text, and
spacers. Text elements may include embedded formulas, rendering results of these formulas
inline.

\subhead{Interactive Controls} Workbook elements may be wired to control 
elements, such as text input, date selectors, or slider controls, enabling the
creation of ``dashboard'' style applications. The controls can be
referenced by column formulas and can be set by parameters to the
Workbook document URL.

\section{Scalability}\label{sec:scalability}
The Workbook implementation has several components to facilitate scalable, interactive collaboration.  We discuss two of them here. 


\subhead{Caching} Workbook employs a hierarchy of caching to reduce
the load on the user's database. An important constraint on our implementation is
that user warehouse data is never stored within the Sigma service cloud. The
first level of caching is within the browser itself. Recent query
results are remembered and re-used, helping the interactivity of
undoing operations or switching to a previous page. The second level
is a directory of recent queries maintained by the Sigma app server.
The directory points to available result sets, stored in the CDW by their query-id, 
which can be re-fetched as requested. It also tracks
in-flight query requests, enabling multiple browsers to share results when
collaboratively editing a document. Finally, the result sets
of user-selected Workbook elements can be materialized into a
warehouse table. The queries for elements that reference the element
are automatically re-written by the Workbook compiler to use these
tables. The materialization can be configured by the user to refresh on
a schedule.

\subhead{In-Browser Evaluation} The browser query-result cache is
augmented with an evaluation engine, written in C++ and compiled to WebAssembly~\cite{haas2017wasm}, which
in many cases can synthesize new results from existing rows already fetched from the CDW.
These local evaluations avoid the latency of a round-trip to the database,
and facilitate the interactivity of workbook editing. 
\cagatay{Can we rephrase the following for clarity?}
In some cases (e.g. lower cardinality tables), we are able to prefetch a resultset that
could be used to fully evaluate all future operations on the table locally in the browser.

\section{Demonstration}\label{sec:demonstration}
We demonstrate the capabilities of Sigma Workbook for common 
business intelligence analysis through 3 example scenarios, highlighting Workbook’s ease of use, scalability, and expressivity. In all the scenarios, we use the On-Time database of the United States domestic airline carrier flights between 1987--2020~\cite{transstats:2020}.
\jlg{This dataset is ~200M rows, which is large but not Very Large. Maybe we can find something bigger that is also interesting?} 
In each scenario, we also show the SQL queries generated by our compiler to produce
these results.

\subhead{Scenario 1: Cohort Analysis} Cohort analysis is a common
analysis with longitudinal datasets. It involves grouping data into
subsets with similar characteristics and comparing how the groups
change over time. The cohort scenario is important enough that earlier research proposed
to extend SQL with new operators to support it~\cite{jiang2016}. Sigma
Workbook has no optimization special to cohort analysis, but it can
be expressed simply with a few basic aggregate expressions. This
analysis is also possible in Power BI but requires comparatively
complex DAX formulas~\cite{power-bi-cohort}.

We demonstrate this analysis in Workbook as follows: 
(1)~Starting with the \sql{FLIGHTS} fact table, we create a self-join
using Workbook's \sling{Rollup} function to identify the date of the first
flight for each plane. This date, truncated to the quarter-year,
identifies the cohort for each plane;
(2)~We then create a hierarchy of grouping levels, first grouping by
cohort and then by flight date truncated by quarter. We compute the total
population of planes in each cohort and, using cross-level references,
the percentage active in each quarter;
(3)~Finally we create a scatter-plot over this dataset, colored by
active population, presenting the synthesized result from over
200M rows of raw data.

\subhead{Scenario 2: Sessionization} Sessionization is an enrichment
where events in time, associated with an entity, are grouped into time
periods known as ``sessions.'' This is useful in marketing, security,
and other applications and is often performed by special-purpose
analysis systems. Relating rows within a SQL database table requires self-joins or
window expressions, both of which are difficult to express in the
language and in many BI systems.

We demonstrate this analysis in Workbook: (1)~Starting with the \sql{FLIGHTS} table, we create a grouping by airplane tail number and then order the base level by flight date.
We infer aircraft servicings from periods of inactivity by adding a window calculation, \sling{Lag} of flight date, and comparing the result with the current flight date. We mark all flights with the
time of service using another window calculation, \sling{FillDown}, as a ``session identifier''; (2)~In a child table element we group first by these discovered sessions and then by cumulative air-time since service was done, and compute cancellation rates for flights at different times in the service life-cycle. We show that users can inspect the rows at each level of aggregation, down to the base; (3)~We visualize this result with a line chart showing how cancellations change with flight hours.

\subhead{Scenario 3: Augmenting Warehouse Data}
Workbook allows users to enrich the shared data of the warehouse with their own data
sources to ``contextualize'' this data.
It enables this in a way familiar to spreadsheet users.

We demonstrate how to use Workbook to augment warehouse data with
external data sources. (1)~First we inspect the FLIGHTS records in
workbook and we discover that they are missing some desired
dimensional data about the airports; (2)~So we perform a web search
and find a plausible dataset that is copied into an editable
Workbook table; (3)~Now we join the new values into the fact table via
a \sling{Lookup} expression; (4)~Upon further inspection we notice the
pasted data is ``dirty'' and correct it with direct editing. We show
that these edits propagate to downstream queries automatically.

\balance
\bibliographystyle{ACM-Reference-Format}
\bibliography{sigma-short}
\end{document}